\documentstyle[preprint,aps]{revtex}
\newcommand{\be}{\begin{equation}} \newcommand{\ee}{\end{equation}} 
\newcommand{\bea}{\begin{eqnarray}}\newcommand{\eea}{\end{eqnarray}}
\def\lb{\label}

\def\ga{\mathrel{\mathchoice {\vcenter{\offinterlineskip\halign{\hfil
$\displaystyle##$\hfil\cr>\cr\sim\cr}}}
{\vcenter{\offinterlineskip\halign{\hfil$\textstyle##$\hfil\cr>\cr\sim\cr}}}
{\vcenter{\offinterlineskip\halign{\hfil$\scriptstyle##$\hfil\cr>\cr\sim\cr}}}
{\vcenter{\offinterlineskip\halign{\hfil$\scriptscriptstyle##$\hfil\cr>\cr\sim\cr}}}}}
\begin{document}
\draft
\title
{\LARGE Deformed strangelets at finite temperature}
\author{Munshi Golam Mustafa{\footnote{Electronic Address: 
mustafa@veccal.ernet.in}}}
\address{Variable Energy Cyclotron center, 1/AF Bidhan Nagar, 
Calcutta 700 064} 
\author{A. Ansari
{\footnote{Electronic Address: ansari@iop.ren.nic.in}}} 
\address
{Institute of Physics, Bhubaneswar 751005, India.}
\date{\today}
\maketitle
\begin{abstract}
\noindent Considering the confinement of massless $u$ and $d$ quarks 
and massive $s$ quarks in an axially symmetric quadrupole shaped 
deformable bag, the stability properties of color-singlet strangelets
at finite temperatures are studied for the baryon number $A \ \leq$ 100.
It is found that, in general, at each $A$ the energy of the  strangelet is
the lowest for the spherical shape. However, it is interesting to find
shell structures for the deformed shapes at some new $A$ values, not found
for the spherical shape. This should  be of relevance for the experimental
search of strangelets. The color projected (singlet) calculation shows
that even for the deformed shapes the shell structures vanish only for
the temperature, $T \ga$ 30 MeV. On the other hand, from a somewhat crude
estimate we also find that at a fixed entropy per baryon 
(instead of a fixed temperature) of the order of unity or higher these
shell structures disappear.
\end{abstract}

\narrowtext

 \vfil
 \eject

\section{Introduction}
\label{int}  

The stability properties of very small strangelets (baryon number $A \
\leq $ 100) were studied recently by Gilson and Jaffe \cite{gj}
and others \cite{fj} in the independent
particle spherical shell model picture where massless $u$ and $d$ quarks,
and massless or massive $s$ quarks are considered to be confined in
a spherical bag with a bag pressure constant, $B$. In view of the hot
environment of ultrarelativistic heavy-ion collisions where strangelets
may be formed, we have recently \cite{mg,mg1} extended the work of Ref.
\cite{gj} to finite temperatures using the method of statistical
mechanics. At a finite temperature ($T > $ \ 0), for a given baryon
number, $A$, a partition function is constructed using the
eigen-energies of the non-interacting ($3A$) quarks confined in the bag.
Then the free energy of the system is minimized with respect to the 
radius of the bag leading to the minimum energy of the system at that
$T$. For $T \ =$ 0 we actually consider $T \ =$ 1 MeV and find
\cite{mg,mg1} that the results are similar to those in Ref. \cite{gj}.
At high temperatures ($T\ga$ 10 MeV) we construct explicitly, following 
projection techniques \cite{mg1,au}, a color-singlet partition function.
At a given $T$ and $A$ the color projection (CP) leads to lowering of
the energy as compared to the unprojected case and when there are shell
closures, the shell pockets become deeper. After CP some of the shell closure
positions also get shifted to other $A$ values. Using $B^{1/4}=145$ MeV
with massless $u$ and $d$ quarks, and massive $s$ quarks ($m_s = 150$
MeV) we found in Ref. \cite{mg1} that in the bulk limit, $A\geq 100$, CP
has little effect and results are similar to those of the liquid drop model
calculations of Jensen and Madsen \cite{jm}. It is also found in Ref.
 \cite{mg1} that for $T\ga
30$ MeV shell structures melt away.

Now, coming back to the main topic of shape deformation, we know that 
many nuclei are 
deformed in the ground state, and this happens simply due to the effect
of shell structures (quantal effect). Therefore, it should be 
worthwhile to investigate the stability properties of strangelets
allowing for the deformation degrees of freedom at $T=0$ as well as at 
$T>0$. Using the same formalism as earlier \cite{mg1} these calculations
can be carried out if the quark eigen-modes in the deformed bag are
known. Actually following the formalism developed by Viollier, Chin
and Kerman \cite{vi} to compute the single-particle energies of quarks
and gluons in a spheroidal bag, we have already made a study \cite{mg2}
of the thermodynamics of a deformed bag with $A=1$ considering two flavors
of quarks $u$ and $d$, and gluons. In the present work the considered
values of temperatures are very low ($T=1$ and 20 MeV) and so like in
Refs. \cite{mg,mg1} here too we have not included gluons in our
calculations.

\section{Formalism}
\label{for}  

The formalism is essentially that of Ref. \cite{mg1} except that now 
the input single-particle energies depend on the quadrupole deformation
parameter which is an additional degree of freedom. The computations
being quite time consuming we have considered only two values of the
temperature $T=1$ and $20$ MeV.

Introducing a deformation parameter, $D$, the surface of a spheroidal bag,
in the body-fixed frame, is given by 
\be
Dx^2 \ + \ Dy^2 \ + \ D^{-2}z^2 \ = \ {R_0}^2 \ , \lb{sur}
\ee
where $R_0$ is the radius of a sphere with the same equivalent volume.
>From Eq. (\ref{sur}) it is clear that the semi-major  and semi-minor axes
of the spheroid are $DR_0$ and $R_0D^{-1/2}$, respectively, with a
volume clearly independent of $D$. For $D$ = 1 the shape is obviously
spherical. A radial coordinate at the surface of the spheroid is given
by 
\be
R(\theta) \ = \ R_0 \left [ a(D) \ + \ {\sqrt{2\over 3}} \ b(D) 
P_2(\cos \theta) \right ]^{-{1/2}} \ , \lb{rad}
\ee
with 
\be
a(D) \ = \ {1\over 3} \left ( 2D \ + \ D^{-2} \right ) \ , \lb{coef1}
\ee
and 
\be
b(D) \ = \ -{\sqrt{2\over 3}} \left ( D \ - \ D^{-2} \right ) \ ,
\lb{coef2}
\ee
where $P_2(\cos\theta)$ is the standard Legendre polynomial of order 2.
The details of the calculations of the eigenmodes for the quarks
confined in the cavity described by Eq. (\ref{sur}) or Eq. (\ref{rad}) are
as given in Refs. \cite{vi,mg2}. The problem is reduced to the 
diagonalization of a real non-symmetric matrix for a fixed parity. The
eigenvalues are functions of $R_0$ and $D$ where $R_0$ enters actually
as $\hbar c/R_0$ with $\hbar c$ = 197.32 MeV fm.

The color-singlet partition function is given by \cite{mg1,au}
\be
{\cal Z}_C(T,R_0,D) \ = \ \int_{\mathrm{SU(3)}} \ {\mathrm d}\mu (g) 
\ {\mathrm{Tr}} \left ( {\hat U}(g) \ e^{-{ {\hat H}/T}} \right ) \ , 
\lb{cp1}
\ee
where ${\hat H}$ denotes the Hamiltonian of the system and 
\be
\int_{\mathrm{SU(N_C)}} \ d\mu (g) \ = \ {1\over {N_C !}} \Big (
\prod^{N_C-1}_{i=1} \int^{\pi}_{-\pi} {{d\theta_i}\over {2\pi}} 
\Big ) \Big [ \prod^{N_C}_{j<k} \big ( 2 \sin {{\theta_j-
\theta_k}\over 2} \big )^2 \Big ]  \ \ , \lb{cp2}
\ee
with $N_C=3$ for SU(3) color group. The chemical potential ($\mu_q=-
\mu_{\bar q}$) dependent partition function can be written as \cite{mg1}
\be
{\cal Z}_C(T,R_0,D,\mu_q) \ = \  \int_{\mathrm {SU(3)}}  d\mu(g) 
\ e^{\Theta} \ , \lb{cp3}
\ee
with
\bea
\Theta &=& {1\over 2} \sum_\alpha \ln\left (
1+{f^q_{\alpha}}^2+2f^q_{\alpha}\cos \theta_1\right ) + {1\over 2}
\sum_\alpha \ln\left ( 1+{f^q_{\alpha}}^2+2f^q_{\alpha}\cos
\theta_2\right ) \nonumber \\
&+& {1\over 2} \sum_\alpha \ln\big [
1+{f^q_{\alpha}}^2+2f^q_{\alpha}\cos (\theta_1+\theta_2) \big ]
+ {1\over 2} \sum_\alpha \ln\left (
1+{f^{\bar q}_{\alpha}}^2+2f^{\bar q}_{\alpha}\cos \theta_1\right )
\nonumber \\
&+& {1\over 2}\sum_\alpha \ln\left ( 1+{f^{\bar
q}_{\alpha}}^2+2f^{\bar q}_{\alpha}\cos \theta_2\right ) + {1\over 2}
\sum_\alpha \ln\left [ 1+{f^{\bar q}_{\alpha}}^2+2f^{\bar
q}_{\alpha}\cos (\theta_1+\theta_2) \right ] \ , \lb{cp4}
\eea
where $f^q_\alpha= e^{- \beta \left (\epsilon^{\alpha}_{q}- 
\mu_q \right )}$ and $f^{\bar q}_\alpha= e^{- \beta \left
(\epsilon^{\alpha}_{\bar q}+\mu_q \right )}$. 

In Eq. (\ref{cp4}) $\beta=1/T$ is the usual inverse temperature with $T$
taken in the units of energy (here MeV) and $\epsilon^\alpha_q$
represents the $R_0$ and $D$ dependent quark(antiquark) single-particle
energies with $\alpha$ denoting the single-particle state labels (spherical
or deformed). The upper limit of the summation
$\alpha$ depends upon the value of $A$ and $T$, such that further
inclusion of states in the summation does not affect the results. In
this sum the summation over quark flavors $u$, $d$, and $s$ is also
understood. The color unprojected (CUP) partition function is simply
given by
\be
{\cal Z}_U(\beta, R_0, D, \mu_q) \ = \ e^{\Theta} \ , \lb{up1}
\ee
where now
\be
\Theta \ = \ \sum_{\alpha} \  \Big [  \ln \big \{ 1 +  \ e^{- \beta
\left ( \epsilon^{\alpha}_{q}- \mu_q \right )} \big \} \ + \ 
\ln \big \{ 1 + \ e^{- \beta \left ( \epsilon^{\alpha}_{\bar q}
+ \mu_q \right )} \big \}  \Big ] \ .  \lb{up2}
\ee

As usual the particle number for a given flavour is calculated from 
\be
N_q \ = \ T {\partial \over {\partial \mu_q}} \left (\ln {\cal Z}\right) 
\ , \lb{num}
\ee
where ${\cal Z}$ is given by Eq. (\ref{cp3}) or Eq. (\ref{up1}), as the
case may be. The baryon number $A$ is fixed by adjusting the value of
the quark chemical potential such that the excess number of $q$ over
${\bar q}$ is $3A$:
\be
(N_u \ - \ N_{\bar u}) \ + \ (N_d \ - \ N_{\bar d}) \ + \ (N_s \ -  
\ N_{\bar s}) \ = \ 3 A \ . \label{qn}
\ee
 The free energy of the system is given by 
\be
F ( T, R_0,D) \ = \ - T\ln{{\cal Z}} \ 
+ \ 3\mu_q A \ + \ BV  \  . \label{f}
\ee
With the above definition of free energy, the condition of stability is
given by
\be
P \ = \ - \left ( {\partial \over {\partial V}} F(T,R_0,D) \right
)_{T,A} = 0\ \ .  \label{p}
\ee
This way at a given $T$ and $D$,  the bag radius parameter $R_0$ is
determined. We may call $R_0$ a radius parameter in the sense that for a
deformed bag it is not actually the radius. In practice we minimize the
free energy with respect to $R_0$, keeping at each step the number
Eq. (\ref{qn}) satisfied. Then the energy of a strangelet is computed
from
\be
E (T, R_0,D) \ = \ T^2{\partial \over {\partial T}} \left 
(\ln{\cal Z}\right ) \ + \ 3\mu_q A \ + \ BV  \ \ , \label{e}
\ee
\noindent where $BV$ is the bag volume energy. 

\section{Results and Discussions}
\label{rd}  

After solving Eqs. (\ref{qn}) and (\ref{p}) selfconsistently for a given
$A$, $T$ and $D$ the energy (mass) of a strangelet is computed according
to Eq. (\ref{e}). As in our earlier works \cite{mg,mg1}, here, too, we have
taken $B^{1/4}$ = 145 MeV, massless $u$ and $d$, and massive  $s$ 
quarks, ($m_s$ = 150 MeV). The variation of the energy per baryon $(E/A)$
as a function of $A$ is plotted in Fig. 1a for $T$ = 1 MeV (practically
the ground state) for spherical and prolate shapes with $D=$ 1.0, 1.2,
1.4, 1.6, 1.8, and 2.0. A similar plot is shown in Fig. 1b for $D$ = 1.0,
0.8 and 0.6; for $D<1$ the shape is oblate. From Eq. (\ref{sur}) it is
clear that the ratio of the semi-major axis $(a)$ to semi-minor axis $(b)$ 
of the spheroid is $D^{3/2}$ from which (just to give an idea of the extent
of deformation) it also follows that approximately 
$a$ = $0.5b$, $2b$ and $3b$ for $D$ = 0.6, 1.6 and 2.0, respectively. These
$a$ and $b$ should not be confused with those defined through Eqs.
(\ref{coef1}) and (\ref{coef2}). From Figs. 1a and 1b it is clear that,
in general, the shell structure is persisting even if the shape of the
bag (strangelet) is deformed, though energitically at every $A$ the
spherical shape leads to the lowest energy configuration. The shell positions
at various baryon numbers ($A_{\mathrm{sh}}$) are shown more explicitly
in Table I. One new thing we find is that with the increase of deformation
(prolate as well as oblate) the shell structure at $A$ = 18 is vanishing 
and, on the contrary, the shell structure at $A$ = 14 is becoming
more pronounced.

In Figs. 2a and 2b we display the variation of the strangeness (actual
value is negative) as a function of $A$ for the prolate and oblate
shapes, respectively at $T$ = 1 MeV. In each figure the $D$ = 1
(spherical) case is presented for the sake of comparison. In the
spherical case we have steplike structure with well-known loading -
unloading characteristics near $A$ = 20 \cite{gj}. With the increase of
deformation the curves are becoming smoother, though still some oscillation
persists, as expected,  at $T$ = 0.

In Fig. 3 we again show the plot of $E/A$ vs $A$ for various prolate
shapes, like in Fig. 1a, but now at $T$ = 20 MeV when the color projected
(singlet) and color unprojected (CUP) curves become distinguishable.
For the CUP case there is hardly any shell structure except near $A$ = 6
for $D$ = 1.0. As can be seen from Table II, this is actually at $A$ = 5
with a small pocket of about 1.0 MeV with the number of $s$-quark,
$n_s$ = 2 (see strangeness vs $A$ plot in Fig. 4). On the other hand,
for CP case and $D$=1.0-1.6, the shell structure appears only at $A$ = 4
with $n_s$ = 0. For $A>$ 4 the strangeness number becomes non-zero, so
much so that even at $A$ = 5 and $D$ = 1, $n_s$=2.3. As Fig. 4 shows,
for $A > $ 10 or so, the $S$ vs $A$ curve becomes almost a straight line
($S/A \sim $ 0.5-0.6) with the strangeness content slightly increasing
with the increas of deformation. One striking thing in Fig. 3 for the CP
case is the development of a shell structure at $A$ = 15 with the
increase of deformation. As can be seen from Table II, this shell pocket
at $A$ = 15 is 0.6 MeV with number of $s$-quark, $n_s \sim$ 7.

>From the discussions so far, it is clear that the color-singlet
constraint leads to more pronounced shell structures. For $T$ = 20 MeV
and $D$=1.0, 1.4 and 2.0 we demonstrate it in Fig. 5 through a plot of a
type of color-corelation energy $\triangle E=
E^{\mathrm{CUP}}/A-E^{\mathrm{CP}}/A$ vs $A$. It clearly shows the
magnified effect of color projection  near the shell closures. In this
figure the first high peak is exactly at $A$ = 4 whereas the second one
is at $A$ = 14 with $A$ = 13 and 15 lying quite nearby. It is
interesting to notice that, though for $D$ = 1.0 there is a clear-cut
shell pocket at $A$ = 14 in the $E/A$ vs $A$ plot, the effect of the
gain in energy due to color projection is getting highly focussed in the
present plot. On the whole for $A>$ 6 the effect of color projection
seems somewhat stronger for $D>1$ but again for $D>1.2$ and $A>20$ it
shows a kind of saturation as far as the large values of $D$ are
concerned.

Finally we want to discuss the effect of deformation vis-a-vis mass of
the $s$-quark. Like in our previous work \cite{mg1}, here, too, we have
considered two values, $m_s$ = 0 and 150 MeV. The $E/A$ vs $A$ plot at
$T$ = 1 MeV is shown in Fig. 6 for $m_s$ = 0 (solid curve) and $m_s$ =
150 MeV (dashed) for $D$ = 1.0 (lower curve), 1.4 (middle) and 2.0
(upper). Again at each $A$ the energy is the lowest for the spherical
shape. The positions and the magnitudes of the shell structures are also
enumerated in Table III for $m_s$ = 0 (see Table I for $m_s$ = 150 MeV).
For $m_s$ = 0 a good (deep) shell structure at $A$ = 6 persists even
with the increase of prolate deformation (oblate not considered) unlike in
the case of massive $s-$quark. Also at $A$ =18 a deeper shell structure
develops with the increase of the prolate deformation in contrast to the
opposite trend for $m_s$ = 150 MeV at the same baryon number, but rather
similar to that at $A$ = 14 when $m_s$ = 150 MeV.

For the $m_s$ = 0 case at $T$ = 20 MeV only the CP case is considered
and the $E/A$ vs $A$ plot for both $m_s$ = 0 (solid) and 150 MeV
(dashed) is displayed in Fig. 7 for $D$ = 1.0 (lower), 1.4 (middle) and
2.0 (upper). For $m_s$ = 150 MeV and $D$ = 1.0 and 1.4 the only shell
structure seen is at $A$ = 4 with $n_s$ = 0, whereas for $D$ = 2.0 a shell
structure has developed at $A$ = 15 (see Table II). It is rather
surprising to see that with $m_s$ = 0 the shell structure at $A$ = 6 is
27.11 MeV, 22.34 MeV and 16.31 MeV deep for  $D$ = 1.0, 1.4 and 2.0,
respectively which are even slightly deeper than that at $T$ = 1 MeV.
Though this feature appears very interesting, it can not be given much
importance as $m_s$ = 0 is unphysical. But it does highlight that the
quantal features of the strangelets, particularly for $A<$ 50, are quite
sensitive to the mass of the $s-$quark.

 Before presenting finally our conclusions in the next section we wish to
make some comments, in view of the work of Ref. \cite{jm}, on the shell
structure features if the E/A vs A plot is made at a fixed entropy per 
baryon ($s$).
>From our color projected numbers at D=1 and $ T = 1, 10, 20, 30 $ MeV we have
computed $s = (E/A - F/A)/T$ at these temperatures. Then a crude graphical
interpolation analysis shows that  for $ s \ga $ 1.0  there appear no shell 
structures. It seems somewhat surprizing but a very interesting  result in 
view of what we have discussed so far at fixed temperatures.

\section{Conclusions}
\label{con}  

We have investigated the stability properties of color-singlet
strangelets at finite temperature within an axially symmetric quadruploe
shape deformable MIT bag model. We find that at each baryon number
$A$($\ \leq$ 100) energetically a spherical shape is preferred at zero
as well as at a finite temperatures. However, with the increase of
deformation we find that the shell closures at some $A$ for the
spherical shape disappear and some new ones appear at different $A$ values.
This should have important implications in the experimental search of
strangelets. The maximum deformed ($D$ = 2.0) prolate shape is
considered such that the ratio of the semi-major to semi-minor axis is about
3:1. On the considered oblate side ($D$ = 0.6) this ratio is 1:2. At a finite
temperature, $T$ = 20 MeV in the CP calculation the shell closure
occures at $A$ = 4 for $D$ = 1.0, 1.2 and 1.6 which is actually a lump
of $u$ and $d$ quarks only.
We may add that this point was missed in our earlier work of Ref.
\cite{mg1}. However, at $D$ = 2.0 and $T$ = 20 MeV a shell structure is
found at $A$ = 15 with $n_s$ = 7.

Considering massless $s$ quarks ($m_s$ = 0), we get, in general, results
similar to as discussed above (see Table III). However, for this case
one interesting result is found that is at $T$ = 20 MeV the shell structure
at $A$ = 6, for $D$ = 1.0, 1.4 and 2.0, is even slightly more pronounced
than that at $T$ = 1 MeV.

It appears that in an  E/A vs A plot at a fixed entropy per baryon ($s$)
there are  no shell structures even for low values of $s \sim $ 1.0 and 
$A \ \leq 30$. A detailed shell model calculation within the present
approach for $ s \ \leq 1.0$ and $A \ \leq 30$ should be interesting
for a proper analysis of the shell structures.

\newpage

\begin{table}
\caption{Shell positions at various baryon numbers ($A_{\mathrm{sh}}$)
and the value of the pocket depth ($E_{\mathrm{sh}}$) defined as
$E_{\mathrm{sh}}=\left( E/A_B\ - \ E/A_{\mathrm{sh}}\right)$ where 
$A_B > A_{\mathrm{sh}}$ and $E/A_B$ has got the highest value. These 
are shown for $T$ = 1 MeV at a few values of the deformation parameter $D$.}

\begin{center}
\begin{tabular}{|l|c|c|c|} 
\hline
& & & \\
$D$ & $A_{\mathrm{sh}}$ & $A_B$ & $E_{\mathrm{sh}}$ (MeV) \\
& & & \\
\hline
  & 6 & 7 & 6.30   \\
  &14 & 15 & 0.17 \\
  &18&22&4.37 \\
1.0&36&40&2.17\\
 &44&47&0.8\\
 &70&82&2.36\\
\hline
  & 6 & 7 & 0.70   \\
  &14 & 16 & 2.24 \\
1.6  &18&19&4.37 \\
&36&41&1.09\\
 &70&83&0.94\\
\hline
  &14 & 16 & 5.00 \\
2.0&36&42&1.34\\
 &72&78&0.40\\
\hline
  &14 & 16 & 3.40 \\
0.6&40&42&0.43\\
 &76&82&0.45\\
\hline
\end{tabular}
\end{center}
\newpage
\caption{Same as Table I for $T$ = 20 MeV. Now color unprojected (CUP)
and color-singlet (CP) both the cases are shown.}
\begin{center}
\begin{tabular}{|l|c|c|c|c|} 
\hline
& & & &\\
$D$ & $A_{\mathrm{sh}}$ & $A_B$ & $E_{\mathrm{sh}}$ (MeV) & Case\\
& & & &\\
\hline
1.0  & 5 & 6 & 0.9&CUP   \\
1.2  &5 & 6 & 0.2&CUP \\
\hline
1.0 & 4 & 7 & 6.0&CP   \\
1.2 & 4 & 6 & 4.7&CP   \\
1.6 & 4 & 6 & 0.2&CP   \\
2.0 & 15 & 17 & 0.6&CP   \\
\hline
\end{tabular}
\end{center}
\newpage
\caption{Same as Table I with $m_s$=0 for CP at $T$ = 20 MeV.}
\begin{center}
\begin{tabular}{|l|c|c|c|} 
\hline
& & & \\
$D$ & $A_{\mathrm{sh}}$ & $A_B$ & $E_{\mathrm{sh}}$ (MeV) \\
& & & \\
\hline
  & 6 & 9 & 26.43   \\
  &18 & 19 & 0.25 \\
  &24&28&3.75 \\
1.0&42&50&4.50\\
 &54&59&1.0\\
 &84&100&4.86\\
\hline
  & 6 & 8 & 18.55   \\
  &18 & 20 & 0.98 \\
1.4  &24&26&0.89 \\
&42&51&3.59\\
 &84&100&4.14\\
\hline
  &6 & 8 & 13.66 \\
2.0&18&21&4.11\\
 &42&51&3.52\\
 &84&100&2.65\\
\hline
\end{tabular}
\end{center}
{$\star$ At $A=$ 18 a shell has developed with the increase of
deformation and has vanished at $A$ = 24 and 54.}
\end{table}

\newpage

\input{pe.ps}

\vspace{0.4in}

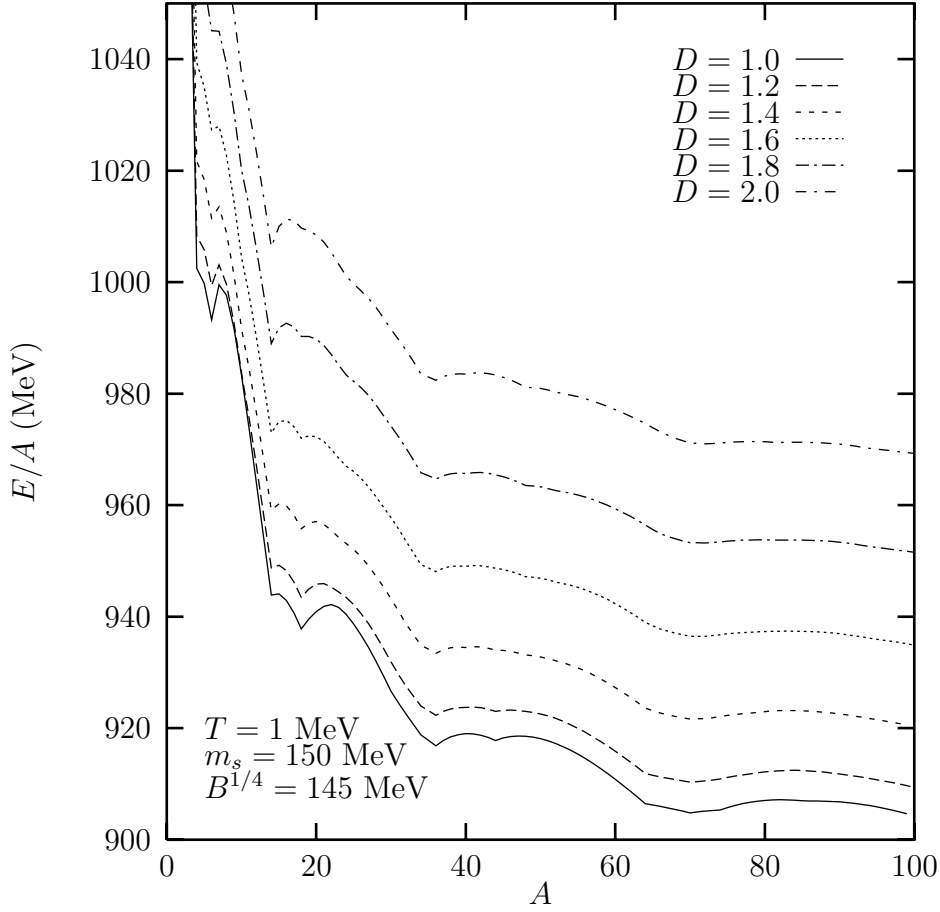
\noindent{ FIG. 1a. { Baryon energy per particle ($E/A$) as a 
fucnction of $A$ for various values of the deformation parameter
($D$=1.0, 1.2, 1.4, 1.6, 1.8 and 2.0) with spherical and prolate shape at  
$T$= 1 MeV. The results are same for both colour projected (CP) and 
unprojected (CUP) cases at $T=$1 MeV. }}

\newpage
\input{oe.ps}

\vspace{0.4in}

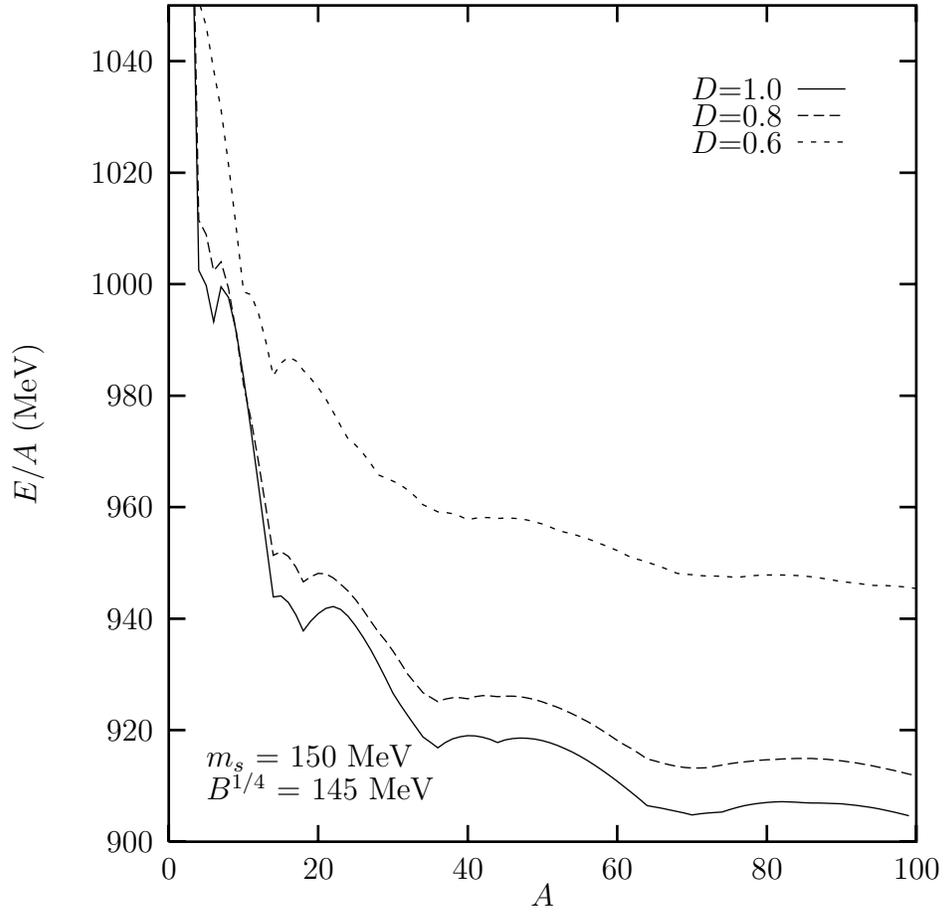
\noindent{ FIG. 1b. Same as FIG. 1a. with spherical and oblate shape for
deformation parameter $D$ = 1.0, 0.8 and 0.6.}
\newpage
\input{dpn.ps}

\vspace{0.4in}

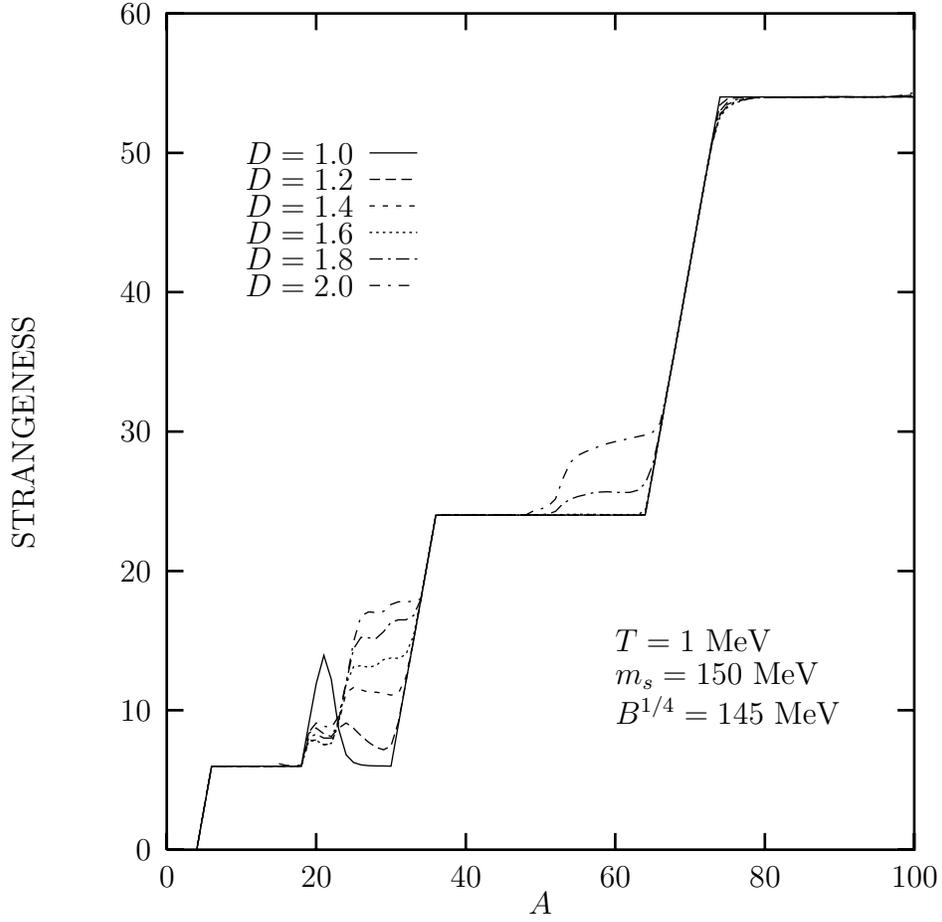
\noindent{FIG. 2a. Strangeness as function of $A$ for different values 
of $D$ (= 1.0, 1.2, 1.4, 1.6, 1.8 and 2.0) at $T$=1 MeV.}
\newpage
\input{don.ps}

\vspace{0.4in}

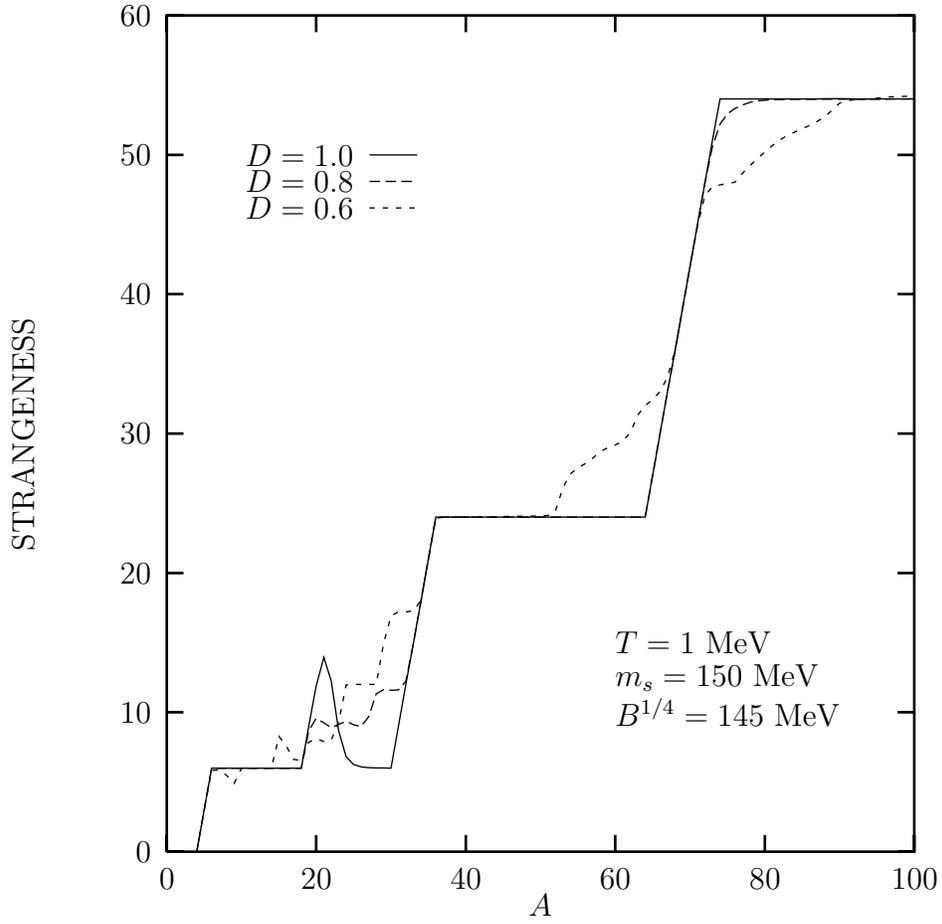
\noindent{ FIG. 2b. Same as FIG. 2a for $D$ = 1.0, 0.8, and 0.6.}
\newpage
\input{pe20.ps}

\vspace{0.4in}

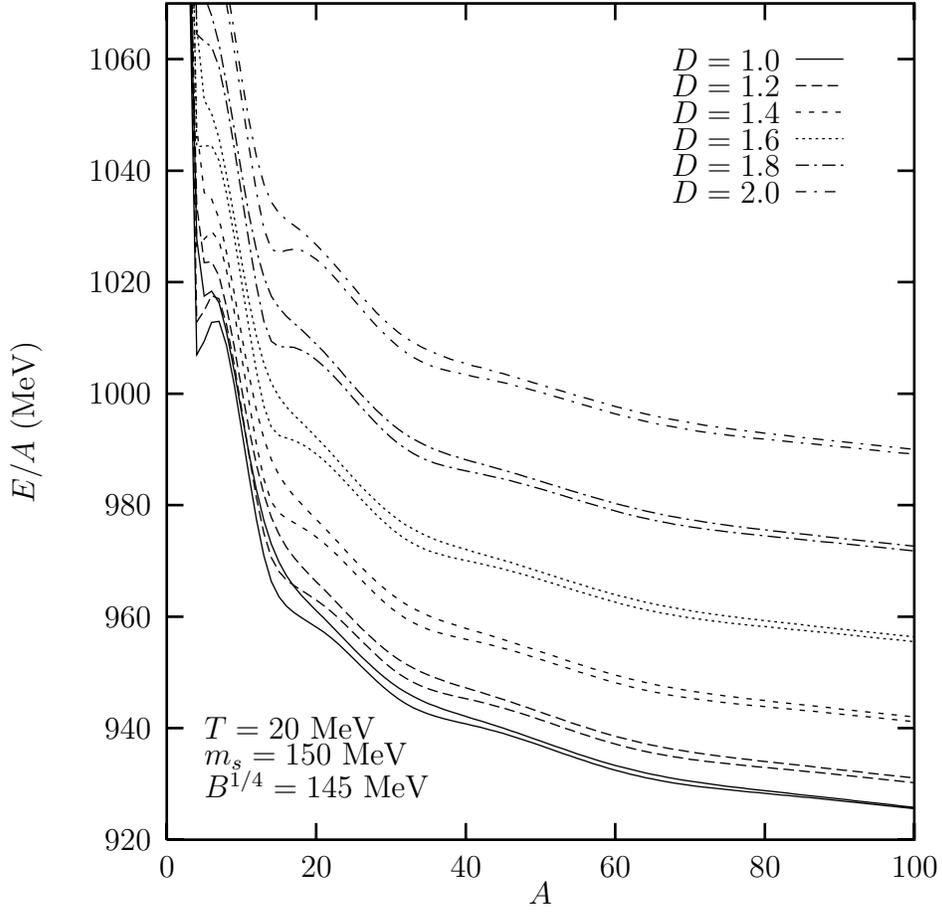
\noindent{ FIG. 3. Same as FIG. 1a with $T=$20 MeV and
deformation values $D=$1.0, $\cdots$ 2.0  for CP and CUP cases.}
\newpage
\input{dpn20_rev.ps}

\vspace{0.4in}

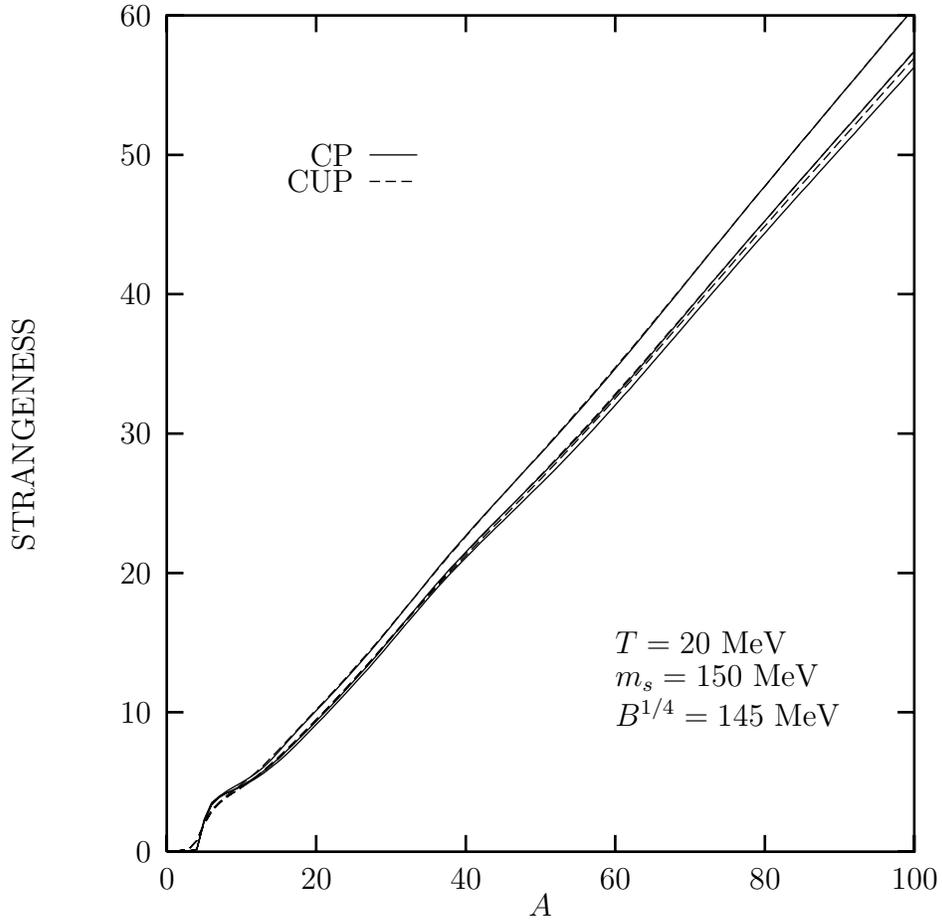
\noindent{FIG. 4. Same as FIG. 2a with $T$=20 MeV for the CP 
(continuous lines) as well as CUP (dashed lines) with $D$ = 1.0, 
1.4 and 2.0 (Down to up).}
\newpage
\input{diff_rev.ps}

\vspace{0.4in}

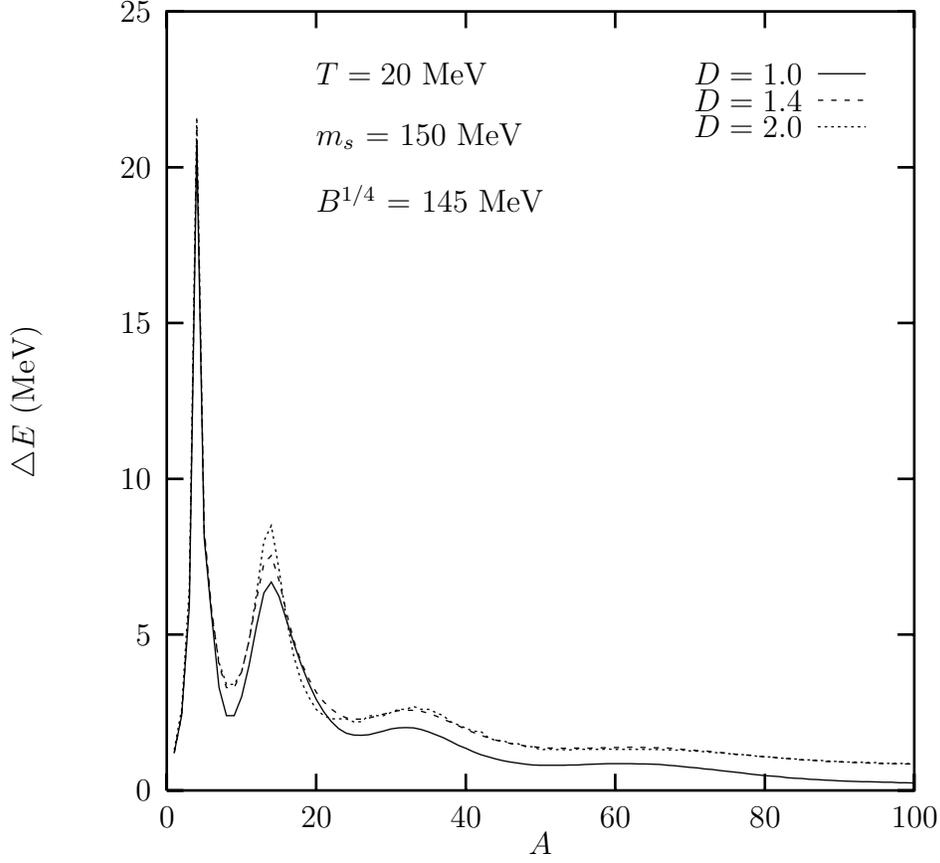
\noindent{FIG. 5. The color-correlation energy  defined as $E^{\mathrm
{CUP}}/A-E^{\mathrm {CP}}/A$  at $T$ = 20 MeV as function of $A$ for 
$D$ =1.0, 1.4 and 2.0.}
\newpage
\input{dem0.ps}

\vspace{0.4in}

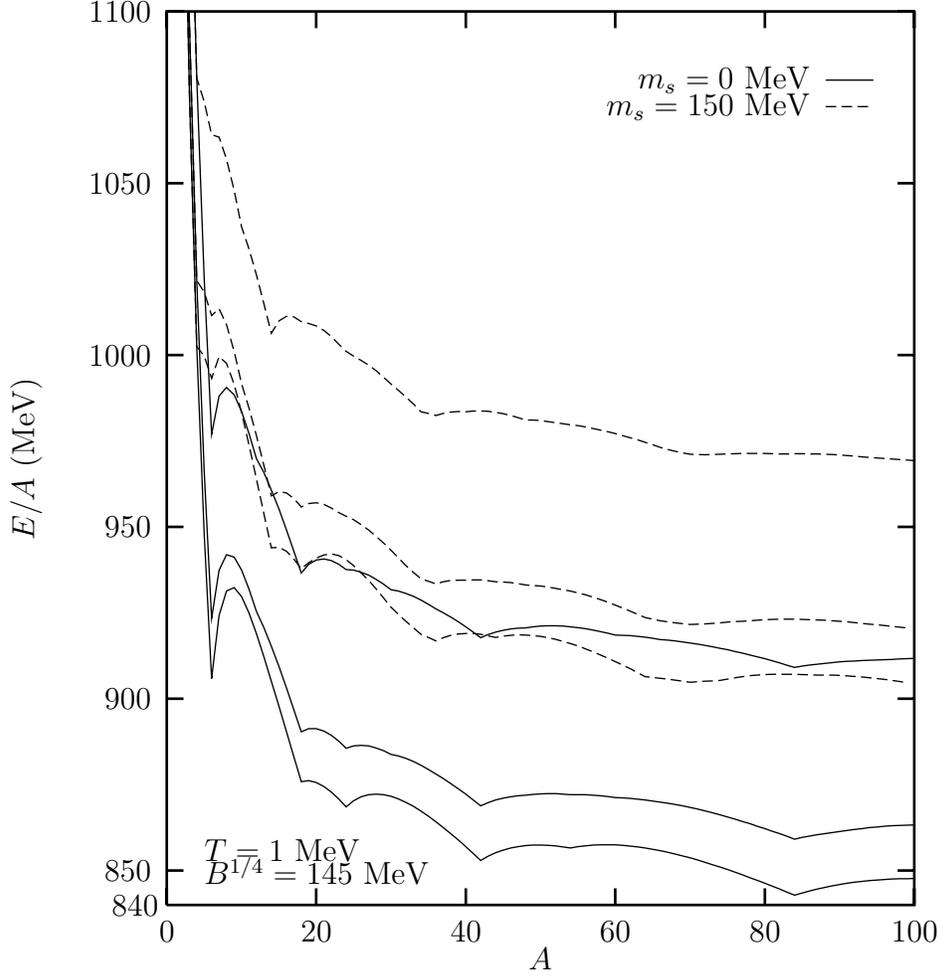
\noindent{FIG. 6. Same as FIG. 1 for $T$ = 1 MeV with $m_s$ = 0
(continuous lines with $D$ = 1 (lower), 1.4 (middle) and 2.0 (upper)) 
and 150 MeV (dashed lines with $D$ =1 (lower), 1.4 (middle) and 2.0 
(upper)) in the CP scheme.}
\newpage
\input{dem20.ps}

\vspace{0.4in}

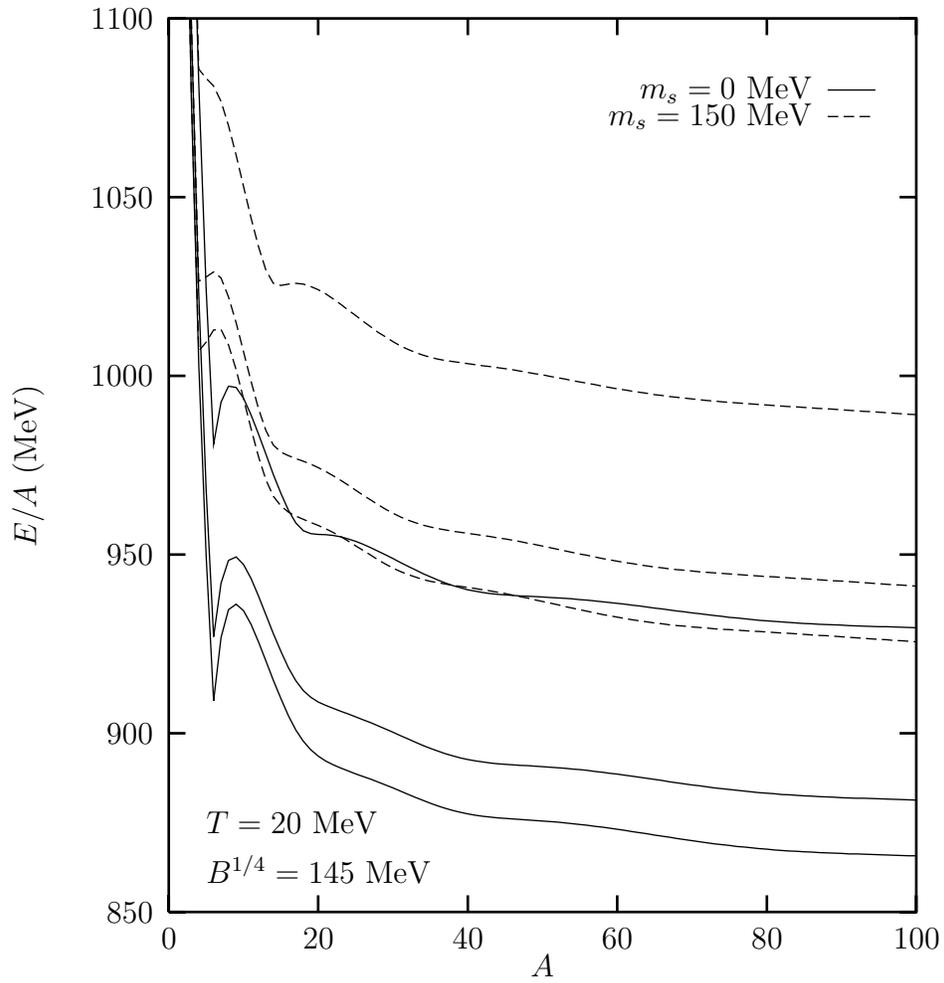
\noindent{FIG. 7.  Same as FIG. 6 with $T$ = 20 MeV.}
\end{document}